# Signal power and energy-per-bit optimization problems in systems mMTC

A. A. Burkov, Assistant Professor, orcid.org/0000-0002-0920-585X, a.burkov@k36.org

Saint-Petersburg State University of Aerospace Instrumentation, 67, B. Morskaia St., 190000, Saint-Petersburg, Russian Federation

**Introduction:** Currently, the issues of the operation of the Internet of Things technology are being actively studied. The operation of a large number of different self-powered sensors is within the framework of a massive machine-type communications scenario using random access methods. Topical issues in this type of communication are: reducing the transmission signal power and increasing the duration of the device by reducing the consumption energy per bit. **The aim of our study:** Formulation and analysis of the tasks of minimizing transmission power and spent energy per bit in systems without retransmissions and with retransmissions to obtain achievability bounds. **Results:** A model of the system is described, within which four problems of minimizing signal power and energy consumption for given parameters (the number of information bits, the spectral efficiency of the system, and the Packet Delivery Ratio) are formulated and described. The numerical results of solving these optimization problems are presented, which make it possible to obtain the achievability bounds for the considered characteristics in systems with and without losses. The lower bounds obtained by the Shannon formula are presented, assuming that the message length is not limited. The results obtained showed that solving the minimization problem with respect to one of the parameters (signal power or consumption energy per bit) does not minimize the second parameter. This difference is most significant for small information message lengths, which corresponds to IoT scenarios. **Practical relevance:** The results obtained allow assessing the potential for minimizing transmission signal power and consumption energy per bit in random multiple access systems with massive machine-type communications scenarios. **Discussion:** The presented problems were solved without taking into account the average delay of message transmission; the introduction of such a limitation should increase the signal power of the transmitted and the consumption energy per bit**.**

## Introduction

In the framework of the currently unfolding 5G communication standard, as well as the development of the next generation 6G, scenarios of the Internet of Things (IoT) are considered [1] [2]. Soon, the number of IoT devices will be on the order of one million per square kilometer, and each of the devices will periodically transmit a small amount of data [3] [4] [5]. Due to the large volume of devices, it is not possible to use scheduling methods to access the channel share. Therefore, within the framework of IoT systems, the use of random access methods is assumed [6] [7][8][9]. With regard to the requirements put forward with the systems (transmission rate, delay, etc.), the following types of IoT are distinguished: massive IoT, critical IoT, broadband IoT and industrial IoT [10]. Massive IoT operates in a massive machine-type communications (mMTC) scenario and describes data acquisition systems with a large number of low-power end devices (such as sensors) that periodically transmit a small amount of data. Examples of the mass Internet of Things are temperature, pressure, light sensors, and meters in smart home technology. The amount of data transferred is small, but the number of IoT devices is very large [11]. The main requirements for this scenario are a large number of devices, stability, low power consumption, delivery of messages with given reliability, as well as a limitation on average delay [12].

The work considers the following scenario. There are a large number of stand-alone user devices and one base station. The level of attenuation in the channel between the base station and all user devices is the same. User devices can transmit data to the base station and receive service messages from the base station. All user devices, at random moments in time, have a small piece of data of the same information length, which the device must transmit to the base station. The total input arrival rate of messages in the system is set. There are two options for this scenario.

*In the first variant*, the system does not provide for the presence of a feedback communication channel. In this case, the user device transmits the message once and deletes it, regardless of the success of the transmission. An important characteristic is the Packet Delivery Ratio (PDR), which can be set in accordance with the requirements of the scenario.

*In the second variant*, the system has a feedback channel, and the user device, according to some algorithm, repeats the transmission until it receives confirmation from the base station that data has been received from this user device.

In what follows, the first variant of the scenario will be called transmission without retransmissions, and the second variant - transmission with retransmissions.

In view of some of the design features of transmission networks, it may be necessary to limit the signal power of data transmission from the user device. This, for example, can be caused by the conditions of the controlling organizations. In this scenario, reducing the transmit signal power is equivalent to decreasing the signal-to-noise ratio to achieve the desired system performance. As noted earlier, in IoT systems, most devices can be powered from an autonomous power source. Therefore, the question arises about increasing the operating time of the device without additional maintenance (replacing the battery or charging). In the described scenario, the increase in the operating time of the user device is reduced to the task of reducing the consumption energy per bit during data transmission.

Accordingly, for both scenarios, 2 minimization problems (task) can be considered.

*First task.* Minimizing signal-to-noise ratio (SNR). In fact, this is the minimization of the transmission signal power for the given parameters: the length of the transmitted message in bits, the intensity of the input arrival rate, packet delivery ratio, and the noise power.

*Second task.* Minimizing consumption energy per bit $\frac{E_b}{N_0}$. In fact, this is the minimization of energy consumption for the transmission of a message with fixed parameters: the length of the transmitted message in bits, the intensity of the input arrival rate, packet delivery ratio, and the noise power spectral density.

It should be noted that when solving the second task within the framework of the feedback system, it is necessary to take into account the number of retransmissions of the message sent by the user device.

As a rule, in works devoted to random multiple access in systems with mMTC, varieties of algorithms such as ALOHA and its modifications are considered [13][14][15]. Within the framework of this work, minimization problems are formulated for *SNR* and $\frac{E_b}{N_0}$ for systems with and without retransmissions, where the optimal ALOHA algorithm will be analyzed as an access algorithm.

The rest of this paper is organized as follows. Section II presents a model of the system, our basic assumptions, and key system characteristics. In Section III, the analysis of the considered scenarios is carried out and the corresponding minimization problems are determined. In addition, in Section IV, several numerical examples of solving the problems posed are proposed, which contain analytical results. Finally, we draw the main conclusions in section V.

## System model

A random multiple access system is considered. Additive White Gaussian Noise (AWGN) is present in the channel. For this system, we introduce a number of assumptions:

*Assumption 1.* The system has one base station and many user devices. The system has an input message arrival rate per unit of time (slot), which is a Poisson distribution with the parameter $\Lambda$ [message / slot]. Each message contains *k* bits of information. Using the modulation and coding scheme (MCS) *A*, on the basis of *k* bits, a signal is generated that contains *n* samples.

Assumption 2. There is a potentially infinite number of user devices (user devices and messages are equal). A user device with a message ready to send is called active.

*Assumption 3.* The user devices and the base station have synchronization, both by samples and by slots. The slot is the time it takes to transmit one message and lasts *n* samples.

*Assumption 4.* We consider a time-discrete communication channel with additive white Gaussian noise (Gaussian Multiple Access Channel, GMAC), defined as $Y = \sum_{i=1}^{K_t} X_i + Z$, where *Y* is the channel output signal, $X_i$ is the signal of the i-th user device, $K_t$ is the number of user devices transmitting a signal per channel in the slot with number *t* (which is a random variable), *Z* is additive white Gaussian noise, and *Z~(0,1)*. The signals transmitted by the user devices contain *n* samples and the maximum energy limitation of each signal $X_i=nSNR$ [16] [17].

The described system model is characterized by a set of parameters: $\Lambda$ [messages/slot], $k$ [bits], $n$ [samples], *SNR* [times] or $\frac{E_b}{N_0}$ [energy/bit]. Additionally, we introduce into consideration the value defined as:

$$\rho \triangleq \frac{\Lambda k}{n}.$$

This value characterizes the average number of bits transmitted per sample. In what follows, it will be called spectral efficiency.

In the next section, taking into account the described model, the analysis of the system, both with guaranteed message delivery and with losses, will be considered. The tasks of minimizing signal power (*SNR*) and consumption energy per bit $\left(\frac{E_b}{N_0}\right)$ will be formulated and described.

## System analysis

First, consider a system operating in without acknowledged mode, with each user device who has a message transmits it at the beginning of the next slot. The number of information bits $k$ to be transmitted to user devices, the PDR $P_d$, and spectral efficiency $\rho$ are indicated.

Add the following assumption to the general model:

*Assumption 5*. The user device transmits a message at the beginning of the next slot, immediately after the message appears, and leaves the system without waiting for confirmation.

The main parameter for this system is the PDR value. There are 2 ways to determine the PDR (by analogy with the definition of the delay in [18]):

*Actual PDR value*. All messages in the system are numbered, and the algorithm works for some time. Then the total number of transmitted messages is divided by the total number of messages that were transmitted during the given period of time.

$$P_{deliv,a} = \lim_{t \to \infty} \frac{N_{deliv}(t)}{N(t)},$$

where $t$ is the operating time of the system, $N_{deliv}(t)$ is the number of messages delivered during time $t$, $N(t)$ is the number of messages sent during time t.

*Virtual PDR value*. A target user device is added to the system at a random moment in time and the probability with which his message will be successfully transmitted is determined.

$$P_{deliv,v} = \lim_{t \to \infty} \Pr \left\{ \begin{array}{l} \text{the packet of the target user added} \\ \text{at time t was successfully deliver} \end{array} \right\}.$$

It was shown in [18] that adding a target user device to the system with a Poisson input stream does not violate the input stream itself and its characteristics. Also in the case of a Poisson input stream, both PDR definitions give the same meaning ($P_{deliv,a}=P_{deliv,v}$).

In this work, to analyze and solve the described minimization problem, the second definition will be used to find the PDR value, and we will denote it as $P_d$.

With this in mind, let's analyze the PDR in general. Let the number of samples $n$, the number of bits transmitted by the user device $k$, the modulation and coding scheme (MCS) $A$, and the SNR value is given. In a random slot, we add a target user device with a message ready for transmission. Then the probability that the message of the target user device will be delivered successfully, in accordance with the second definition of PDR, is calculated as follows:

$$P_d = \Pr \left\{ \begin{array}{l} 0 \text{ user devices} \\ \text{appeared in the system} \end{array} \right\} \Pr \left\{ \begin{array}{l} \text{message decoded} \\ \text{successfully} \end{array} \right\}.$$

Using the formula for the probability of occurrence of $i$ messages with a Poisson input stream with a parameter $\Lambda$, we can write:

$$\Pr \left\{ \begin{array}{l} i \text{ user devices} \\ \text{appeared in the system} \end{array} \right\} = \frac{(\Lambda)^i}{i!} e^{-\Lambda}.$$

Let there be some function $P_e(A,n,k,SNR)$ that allows you to determine the probability of decoding error for given parameters $A$, SNR, $n$ and $k$, then:

$$\Pr \left\{ \begin{array}{l} \text{message decoded} \\ \text{successfully} \end{array} \right\} = 1 - P_e(A,n,k,\text{SNR}).$$

Then, taking into account the fact that $\Lambda = \dfrac{\lambda n}{k}$, we get the expression:

$$P_d = e^{-\rho \frac{n}{k}} \left(1 - P_e(A,n,k,SNR)\right). \tag{1}$$

Polyansky's formula [16] $p_e(n,k,\text{SNR}) = Q\left(\dfrac{n\dfrac{1}{2}\log_2(1+\text{SNR}) + \dfrac{1}{2}\log_2 n - k}{\sqrt{n\dfrac{\text{SNR}}{2}\dfrac{\text{SNR}+2}{(\text{SNR}+1)^2}}\log_2 e}\right)$ allows determining the probability of decoding error for given parameters SNR, $n$ and $k$. Then we define the probability of successful decoding as 1- $p_e(A,n,k,SNR)$. Substitute in formula (1) and get:

$$P_d = e^{-\rho \frac{n}{k}}\left(1 - p_e(k,n,SNR)\right). \tag{2}$$

Then the *first minimization task* can be formulated:

$$\begin{aligned}&\text{given}: k, \rho, P_d\\&\text{minimaze}: SNR \text{ in } n\\&\text{subject to}:\\&P_d = e^{-\rho \frac{n}{k}}\left(1 - p_e(k,n,SNR)\right)\end{aligned} \tag{3}$$

The solution to this task will be the minimum SNR value at which the required PDR is achieved for a given number of transmitted bits $k$ and spectral efficiency $\rho$.

Consider the formulation of a similar task but from the point of view of minimizing the expended energy per bit. It is known that SNR and $\dfrac{E_b}{N_0}$ are related through the following expression:

$$\frac{E_b}{N_0} = \frac{n\,SNR}{2k}. \tag{4}$$

With this in mind, expression (2) can be written as:

$$P_d = e^{-\rho \frac{n}{k}}\left(1 - p_e\left(k,n,\frac{2k}{n}\frac{E_b}{N_0}\right)\right). \tag{5}$$

Then the *second minimization task* (consumption energy per bit, for a given number of transmitted bits and spectral efficiency) can be formulated as follows:

$$\begin{aligned}&\text{given}: k, \rho, P_d\\&\text{minimaze}: \frac{E_b}{N_0} \text{ in } n\\&\text{subject to}:\\&P_d = e^{-\rho \frac{n}{k}}\left(1 - p_e\left(k,n,\frac{2k}{n}\frac{E_b}{N_0}\right)\right)\end{aligned} \tag{6}$$

Consider the second system, in which the user device transmits his message according to some algorithm until it is successfully delivered. The number of information bits $k$ and the spectral efficiency $\rho$ are given.

It is necessary to change the model of the system by adding an assumption to it:

*Assumption 5* [*]. By the beginning of each slot, all user devices and BS know how many user devices have a message ready for transmission $M_a$. Each user device transmits his message with probability $p = \dfrac{G}{M_a}$, where $G$ is the algorithm parameter ($0 < G \leq 1$), until it is successfully delivered.

The critical input arrival rate can be limited by the following inequality:

$$\Lambda \leq \Pr \left\{ \begin{array}{l} \text{probability of transmission} \\ \text{by } \textit{one} \text{ user device} \end{array} \middle| G \right\} \Pr \left\{ \begin{array}{l} \text{message decoded} \\ \text{successfully} \end{array} \right\}. \qquad (7)$$

Then, taking into account the spectral efficiency [bit/sample] and by analogy with the system without retransmissions, the expression can be written as:

$$\rho = \frac{k}{n} G e^{-G} \left( 1 - p_e(k, n, SNR) \right). \qquad (8)$$

It is worth noting that for the algorithm without retransmissions, taking into account SNR, one can find the value of $G$ that maximizes the spectral efficiency. To do this, you need to solve the following optimization problem:

$$G_{opt} = \max_{G} G e^{-G}. \qquad (9)$$

It is well known that the solution to this optimization problem is $G=1$ [19].

Then the third minimization task can be formulated:

$$\begin{array}{l} \text{given}: k, \rho \\ \text{minimaze}: SNR \text{ in } n \\ \text{subject to:} \\ \rho = \dfrac{k}{n} e^{-1} \left( 1 - p_e(k, n, SNR) \right) \end{array} \qquad (10)$$

Consider the same system, but from the point of view of minimizing consumption energy per bit. In contrast to solving the problem in terms of signal power, it is necessary to take into account the average number of messages $S$ sent by a user device. From work [20] it follows that the average number of transmissions is determined by the expression:

$$S = \frac{kG(1 - \pi_0)}{\rho n}, \qquad (11)$$

where $S$ is the average number of messages sent by the user device, $k$ is the number of information symbols, $n$ is the number of samples, $\rho$ is the spectral efficiency, $\pi_0$ is the stationary probability that there are no active user devices in the system.

Considering that the value $\rho$ specified in the solution of the task is critical, then the value $\pi_0 \to 0$. Then, taking into account the number of transmissions, the SNR values and $\dfrac{E_b}{N_0}$ are related by the following expression:

$$\frac{E_b}{N_0} = \frac{Sn\,SNR}{2k} = 2\frac{E_b}{N_0}\frac{\rho}{G}. \qquad (12)$$

In accordance with expression (12) and task (10), we can formulate the fourth *minimization task* (for $\dfrac{E_b}{N_0}$):

$$\begin{aligned}
&\text{given}: k, \rho \\
&\text{minimaze}: \frac{E_b}{N_0} \text{ in } n, G \\
&\text{subject to:} \\
&\rho = \frac{k}{n} G e^{-G}\left(1 - p_e\left(k, n, 2\frac{E_b}{N_0}\frac{\rho}{G}\right)\right)
\end{aligned} \qquad (13)$$

All considered minimization tasks are presented in Table 1.

*Table 1. Formulation of minimization tasks*

*Таблица 1. Формулировки задач минимизации.*

|  | Minimizing signal power (SNR) | Minimization of energy per bit costs $\left(\dfrac{E_b}{N_0}\right)$ |
|---|---|---|
| A system without retransmissions | **Task 1**<br>given: $k, \rho, P_d$<br>minimaze: $SNR$ in $n$<br>subject to:<br>$P_d = e^{-\rho\frac{n}{k}}\left(1 - p_e(k, n, SNR)\right)$ | **Task 2**<br>given: $k, \rho, P_d$<br>minimaze: $\dfrac{E_b}{N_0}$ in $n$<br>subject to:<br>$P_d = e^{-\rho\frac{n}{k}}\left(1 - p_e\left(k, n, \dfrac{2k}{n}\dfrac{E_b}{N_0}\right)\right)$ |
| A system with retransmissions | **Task 3** | **Task 4** |
| | given: k, ρ <br> minimaze: $SNR$ in $n$ <br> subject to: <br> $\rho = \dfrac{k}{n} e^{-1}\left(1 - p_e(k, n, SNR)\right)$ | given: k, ρ <br> minimaze: $\dfrac{E_b}{N_0}$ in $n, G$ <br> subject to: <br> $\rho = \dfrac{k}{n} G e^{-G}\left(1 - p_e\left(k, n, 2\dfrac{E_b}{N_0}\dfrac{\rho}{G}\right)\right)$ |

The solutions to these tasks allow us to determine the achievability bounds of signal power and consumption energy per bit in random multiple access systems.

If we consider the solution of the presented tasks for $k \to \infty$, $n \to \infty$ and the speed $R=k/n$, taking into account the Shannon theorem for discrete channels, then we can obtain the lower bounds presented in Table 2.

*Table 2. Lower bounds for solving minimization tasks*

*Таблица 1. Нижние границы решения задач минимизации*

| | Minimizing signal power (SNR) | Minimization of energy per bit costs $\left(\dfrac{E_b}{N_0}\right)$ |
|---|---|---|
| A system without retransmissions | **Task 1** <br><br> $SNR > 2^{-\dfrac{2\rho}{\ln(P_d)}} - 1$ | **Task 2** <br><br> $\dfrac{E_b}{N_0} > \dfrac{2^{-\dfrac{2\rho}{\ln(P_d)}} - 1}{-\dfrac{2\rho}{\ln(P_d)}}$ |
| A system with retransmissions | **Task 3** <br><br> $SNR > 2^{\dfrac{2\rho}{e^{-1}}} - 1$ | **Task 4** <br><br> $\dfrac{E_b}{N_0} > \dfrac{\left(2^{\dfrac{2\rho}{Ge^{-G}}} - 1\right)G}{2\rho}$ |

In the next section, the numerical results of solving these tasks will be presented and qualitative analysis of the results will be carried out.

**Numerical results**

Figures 1 and 2 show the results of solving minimization tasks for a system without retransmissions (tasks 1 and 2), for a given PDR $P_d=0.9$ at spectral efficiency $\rho=0.1$ and

ρ=*0.12*. The values marked "converted" mean that the minimization task was solved by one of the parameters and was recalculated for the second parameter in accordance with formula (4). It should be noted that in the case of an increase in the required PDR, the obtained values of the achievability bounds increase. It also follows from the graphs that the solution of the problem of minimization in terms of signal power does not give a solution minimization in terms of consumption energy per bit and vice versa. As *k* grows, these differences decrease. The lower bounds are obtained in accordance with the expressions from Table 2. As can be seen from the figure, with an increase in the number of information bits, the achievability bound tends to the lower bound. As *k*→∞, the results will be the same. However, the Internet of Things has short message lengths. The tasks posed earlier can be changed by minimizing one of the parameters with a constraint on the other, but such solutions will lie above the obtained bounds.

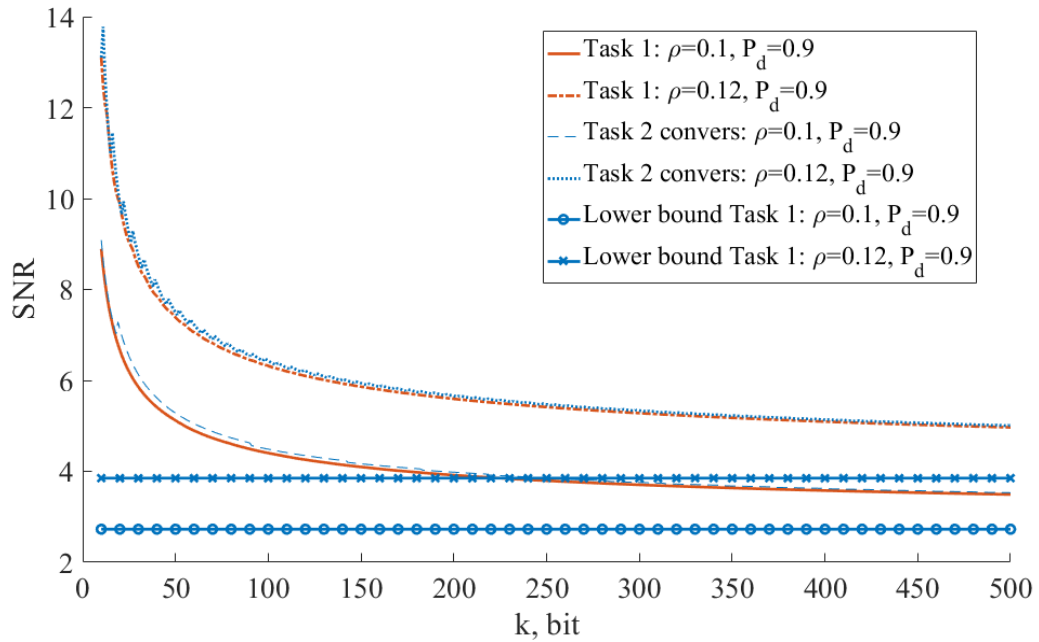

Figure 1. Achievability bounds and lower bounds for SNR from *k* for tasks 1 and 2 at $P_d=0.9$.

Рисунок 1. Граница достижимости и нижняя граница для SNR от *k* для 1 и 2 задач при $P_d=0.9$.

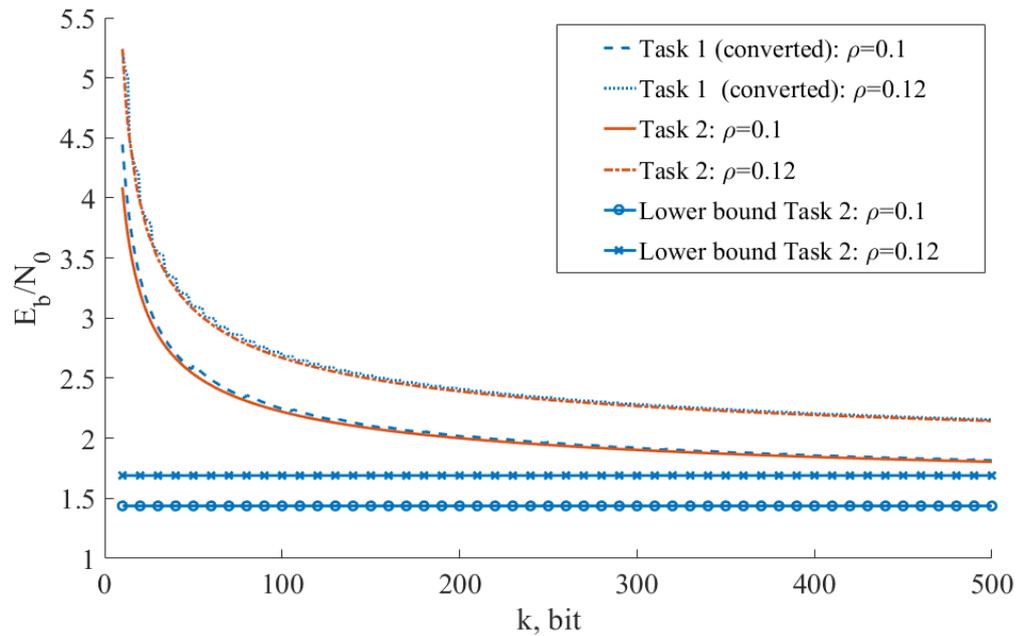

Figure 2. Achievability bounds and lower bounds for $\frac{E_b}{N_0}$ from $k$ for tasks 1 and 2 at $P_d=0.9$.

Рисунок 2. Граница достижимости и нижняя граница для $\frac{E_b}{N_0}$ от $k$ для 1 и 2 задач при $P_d=0.9$.

Figures 3 and 4 show the results of solving the minimization problems for all tasks, for a given number of information bits $k$ equal to 50. The PDR for the system without repetitions is *0.65* and was chosen for the convenience of the scale on the graph. The values marked "converted" also mean that the minimization task was solved by one of the parameters and was recalculated for the second parameter in accordance with formulas (4) - for tasks 1 and 2, (12) - for tasks 3 and 4. As noted earlier, solving the task for the system with retransmissions by SNR, the optimal system parameter $G$ is equal to 1. However, when solving this task by $\frac{E_b}{N_0}$, the values of this parameter turn out to be below 1 and tend to this value with increasing spectral efficiency. By optimizing this parameter, there is a gain in consumption energy per bit in systems with retransmissions. If this parameter is equal to 1 in both tasks, they give similar results.

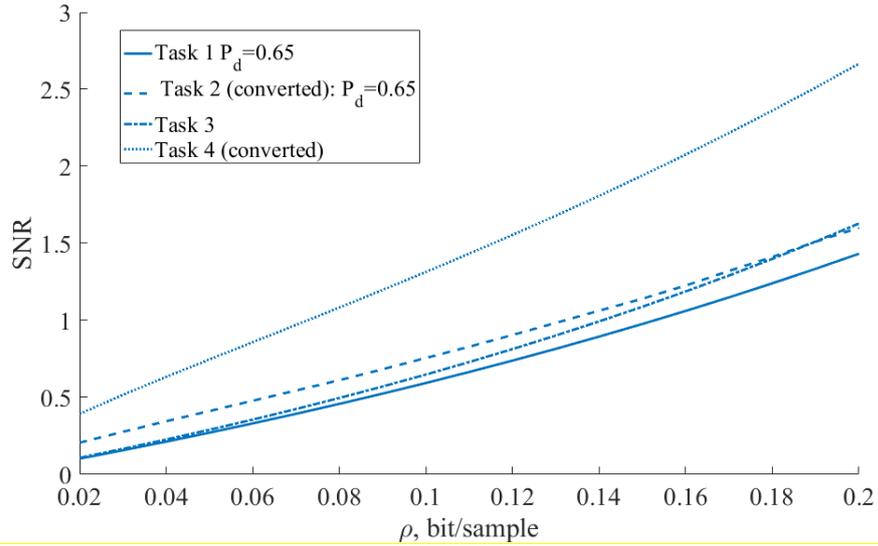

Figure 3. Achievability bounds SNR from all tasks for ρ at $k = 50$.

Рисунок 3. Границы достижимости для SNR от ρ для всех задач при $k$=50.

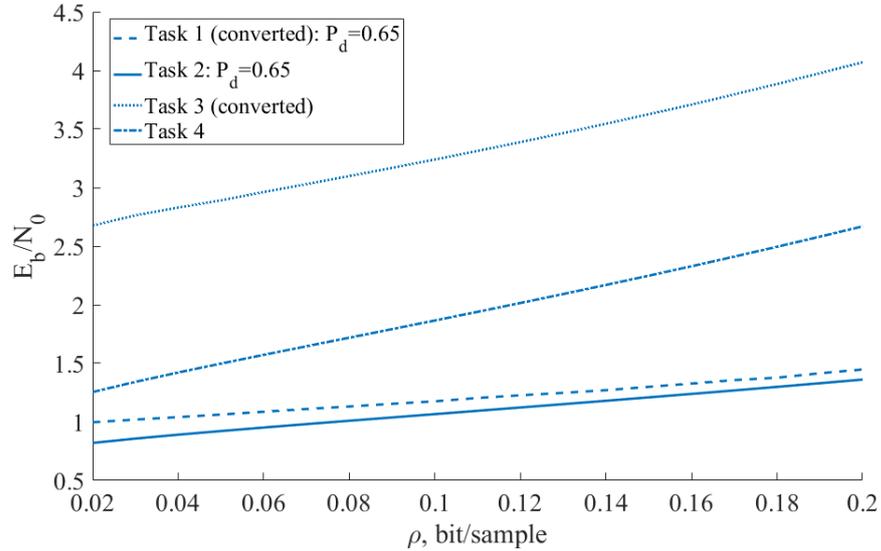

Figure 4. Achievability bounds $\frac{E_b}{N_0}$ from all tasks for ρ at $k = 50$.

Рисунок 4. Границы достижимости для $\frac{E_b}{N_0}$ от ρ для всех задач при $k$=50.

## Conclusion

The paper describes a model of a system with a potentially unlimited number of user devices in a Gaussian MAC. Within the framework of this model, systems without retransmissions and with retransmissions are considered. Four tasks of minimizing signal power and consumption energy per bit for given system parameters were formulated and described: the number of information bits, the spectral efficiency of the system, and the packet delivery ratio. For the problems formulated, the lower bounds obtained by the Shannon formula are presented under the assumption that the message length is not limited. With an increase in the number of information bits, the results of solving tasks and the corresponding achievability bounds will tend to lower bounds. Numerical solutions for the assigned tasks are presented. The results obtained showed that solving the minimization problem with respect to one of the parameters (signal power or consumption energy per bit) does not minimize the second parameter. This difference is most significant for small information message lengths, which corresponds to IoT scenarios. Difference a dB at typical value of information bits k = 50 bits and spectral efficiency *ρ=0.1* according to SNR: for tasks 1 and 2 with a PDR=0.9 is 0.13dB; for tasks 3 and 4 is 3.07dB. Results for $\frac{E_b}{N_0}$ in similar conditions: for tasks 1 and 2 with a PDR=0.9 is 0.12dB; for tasks 3 and 4 is 2.4dB. However, in IoT scenarios, the number of information bits is assumed to be small. For a system with retransmissions, with minimization of energy per bit, the optimal transmission parameter *G* is less than 1 and tends to it with an increase in the required spectral efficiency. When minimizing signal power in such a system, this parameter is always 1.

The results obtained allow us to assess the potential for minimizing transmission signal power and consumption energy per bit in random multiple access systems in the framework of IoT scenarios. Sometimes, in practice, the transmitter signal power can be limited and it is required to minimize the energy consumption, therefore, one of the parameters can be minimized with a restriction on the other, then the boundaries obtained in the work will be a lower estimate for solving such problems.

## Acknowledgment

The research was supported by RFBR, project number 19-37-90041.

**Литература**